\documentstyle[aaspp4,12pt,flushrt]{article}

\newcommand{\kms}{km~s$^{-1}$}

\newcommand{\br}{\hbox{$B\!-\!R$}}
\newcommand{\ha}{\hbox{{\rm H}$\alpha$}}		

\begin{document}
\title{An Imaging and Spectroscopic Survey of
Galaxies within Prominent Nearby Voids II.
Morphologies, Star Formation, and Faint Companions}

\author{Norman A. Grogin and Margaret J. Geller} 

\affil{Harvard-Smithsonian Center for Astrophysics,
60 Garden Street, Cambridge, MA 02138\\ 
E-mail: ngrogin, mgeller@cfa.harvard.edu
}

\slugcomment{To appear in the {\it Astronomical Journal}, Jan.~2000}

\begin{abstract}
We analyse the optical properties of
$\sim\!300$ galaxies within and around three prominent voids of the
Center for Astrophysics Redshift Survey (\cite{gh89}).  We determine CCD
morphologies and \ha\ equivalent widths from our imaging and
spectroscopic survey (Grogin \& Geller 1999).  We also describe a 
redshift survey of 250 neighboring galaxies in the imaging survey fields.  We
assess the morphology-density relation, EW(\ha)-density relation, and the
effects of nearby companions for galaxies in low-density environments
selected with a smoothed large-scale ($5h^{-1}$~Mpc) galaxy number
density $n$.

Both the morphological mix and the \ha\ linewidth distribution of
galaxies at modest underdensites, $0.5 < (n/\bar n) \leq 1$, are
indistinguishable from our control sample at modest overdensities, $1
< (n/\bar n) \leq 2$.  Both density regions contain a similar fraction
of galaxies with early-type (E and S0) morphologies and with
absorption-line spectra ($\approx\!35$\%).  At the lowest densities,
$(n/\bar n) \leq 0.5$, there is a $3\sigma$ shift in the distribution
of EW(\ha) away from absorption-line systems (only $\approx\!15$\%)
and toward emission-line systems with active star formation ---
EW(\ha)$\sim 40$\AA--100\AA.  There is a $2\sigma$ shift in
the morphological distribution away from early-types and toward
irregular and peculiar morphologies.

The redshift survey of projected companions, 80\% complete to $m_R =
16.13$, demonstrates that the incidence of a close companion in
redshift space is insensitive to global density over the range
we investigate ($0.16 < n/\bar n \leq 2$).  However, the typical
velocity separation of close pairs drops significantly ($>3\sigma$)
from $\Delta cz\gtrsim 200$~\kms\ at $0.5 < n/\bar n \leq 2$ down to
$\Delta cz = 103\pm20$~\kms\ at $n \leq 0.5\bar n$.  In the
lowest-density environments, galaxies with companions clearly
($\sim\!4\sigma$) have stronger star formation than comparable galaxies at
larger global density ($0.5 < n/\bar n \leq 2$).  On the other hand,
the distribution of EW(\ha) for galaxies {\sl without} nearby
companions (closer than $\approx 150h^{-1}$~kpc and 1000~\kms) varies
little over the entire density range.

These results, combined with the luminosity- and color-density
relations of this sample (Grogin \& Geller 1999), suggest that the
formation and evolution of field galaxies are insensitive to
large-scale underdensity down to a threshold of roughly half the mean
density.  The differences in galaxy properties at the lowest global
densities we can explore ($n\leq 0.5\bar n$) may be explained by: 1) a
relative scarcity of the small-scale primordial density enhancements
needed to form massive early-type/absorption-line galaxies; and 2)
present-day galaxy encounters which are relatively more effective
because of the lower velocity dispersion on small scales ($\lesssim
200h^{-1}$~kpc) we observe in these regions.  In the voids, where the
luminous galaxies presumably formed more recently, there should be
more gas and dust present for active star formation triggered by
nearby companions.  
\end{abstract}

\begin{figure}[bp]
{
\plotone{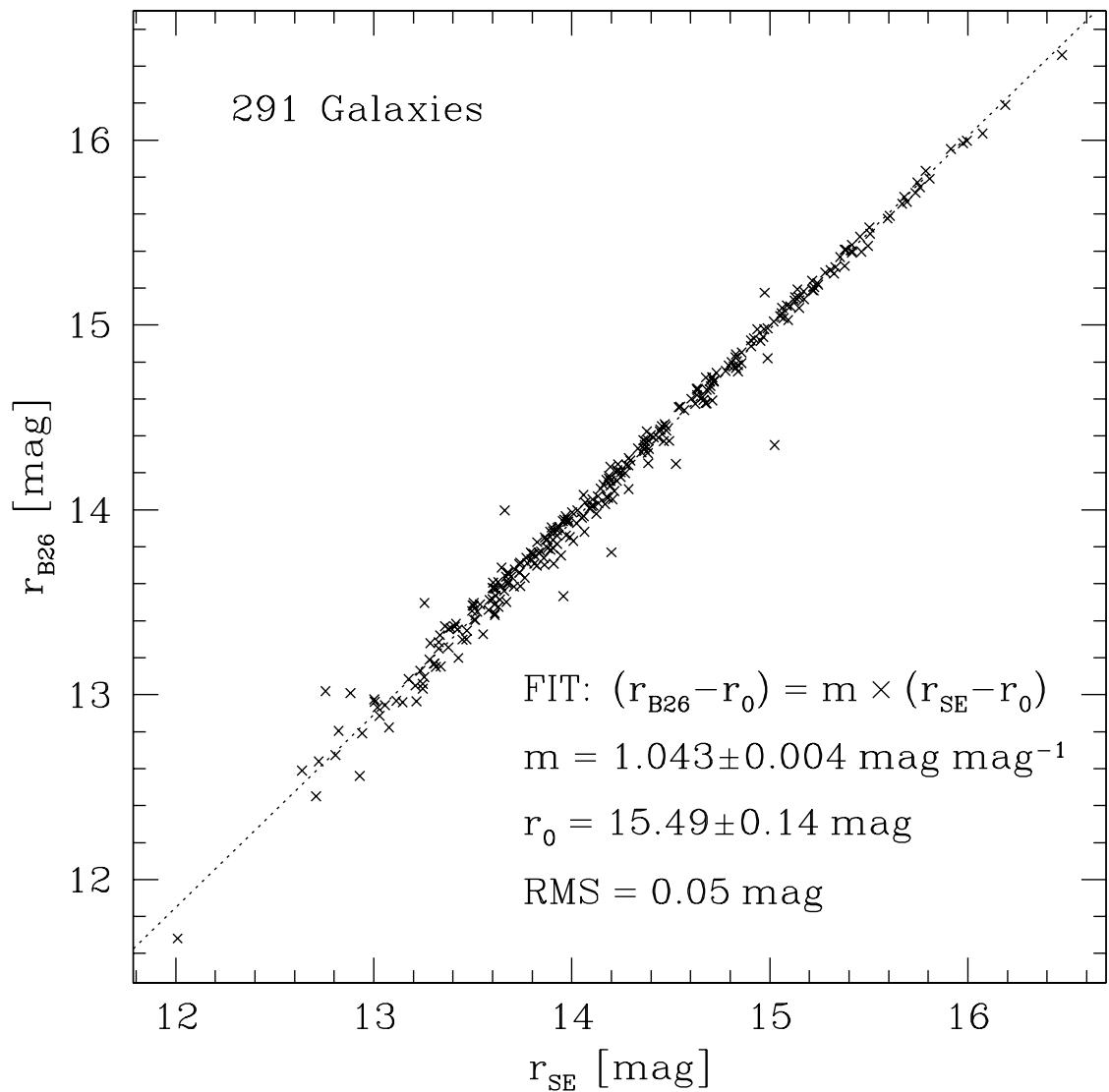}
\figcaption[Grogin.fig1.ps]{
SExtractor $R$-band magnitude, $r_{\rm SE}$, compared
with the $R$-band magnitude at the $26\mu_B$ isophote, $r_{B26}$, for 291
galaxies (Tab.~\ref{brighttab}).  The dotted line shows the
least-squares fit with $2\sigma$-clipping.  The parameters of the linear
fit and their errors are listed at lower right.
\label{isosex}}
}
\end{figure}

\begin{section}{Introduction}
In the last decade, wide-angle redshift surveys have revealed
large-scale structure in the local universe comprised of coherent
sheets of galaxies with embedded galaxy clusters, bounding vast
($10^{5-6}$ Mpc$^3$) and well-defined ``voids'' where galaxies are
largely absent.  The influence of these structures' large-scale
density environment upon galaxy properties has been a continuing
source of debate and is of interest for constraining proposed models
of galaxy formation and evolution.  The morphology-density relation
(e.g., \cite{dr80,pg84}), which quantifies the increasing fraction of
ellipticals and lenticulars with local density, is one of the most
obvious indications of environmental dependence for densities greater
than the mean.  In the lowest density regions, the voids, the
observational evidence of trends in morphological mix, luminosity
distribution, star formation rate, etc., is still rudimentary because
of the intrinsic scarcity of void galaxies and the difficulties in
defining an unbiased sample for study.  Here we use a broadband
imaging and spectroscopic survey of a large optically-selected sample
to compare the properties of galaxies in voids with their counterparts
in denser regions.

We refer the reader to the first paper of this study (Grogin \& Geller
1999, hereafter \cite{pap1}) for a more detailed review of the
previous theoretical and observational research into void galaxies,
which we summarize here.  Proposed theories of galaxy formation and
evolution have variously predicted that the voids contain ``failed
galaxies'' identified as diffuse dwarfs and Malin 1-type giants
(\cite{ds86,hsw92}), or galaxies with the same morphological mix as
higher-density regions outside clusters (\cite{bss98}), or no luminous
galaxies at all for the want of tidal interactions to trigger star
formation (\cite{lac93}).  Some of these theories have already met
serious challenge from observations which have shown that dwarf
galaxies trace the distribution of the more luminous galaxies and do
not fill the voids (\cite{khe97,phe97,bing89}), and that the low
surface-brightness giants are relatively rare and not found in the
voids (\cite{sz96b,bot93,wei91,hk89}).

Most previous studies of void galaxies have focused on emission
line-selected and $IRAS$-selected objects in the Bo\"otes void at
$z\sim0.05$ (\cite{koss1,koss2}); all have been limited to a few dozen
objects or fewer.  The galaxies observed in the Bo\"otes void: 1) are
brighter on average than emission-line galaxies (ELGs) at similar
redshift and contain a large fraction ($\approx40$\%) with unusual or
disturbed morphology (\cite{cwh97}); 2) have star formation rates
ranging from 3--55 $\cal M_\odot$ yr${}^{-1}$, up to almost three
times the rate found in normal field disk systems (\cite{w95}), in
apparent contrast to the Lacey et al.\ (1993) model prediction; and 3)
are mostly late-type gas-rich systems with optical and H{\sc i} properties
and local environments similar to field galaxies of the same
morphological type (\cite{sz96a,sz96b}).  

Szomoru et al.~(1996) conclude that the Bo\"otes void galaxies formed
as normal field galaxies in local density enhancements within the
void, and that the surrounding global underdensity is irrelevant to
the formation and evolution of these galaxies.  Because the Bo\"otes
void galaxies are not optically-selected, though, their properties may
not be representative of the overall void galaxy population.  On the
other hand, similar conclusions were drawn by Thorstensen et
al.~(1995) in a study of 27 Zwicky Catalog (\cite{cgcg}, hereafter
CGCG) galaxies within a closer void mapped out by the Center for
Astrophysics Redshift Survey (\cite{gh89}; hereafter CfA2).  The
fraction of absorption-line galaxies in their optically-selected
sample was typical of regions outside cluster cores, and the {\sl
local} morphology-density relation appeared to hold even within the
global underdensity.

Our goal is to clarify the properties of void galaxies by collecting
high-quality optical data for a large sample with well-defined
selection criteria.  We thus obtained multi-color CCD images and high
signal-to-noise spectra for $\sim\!150$ optically-selected galaxies
within prominent nearby voids.  We work from the CfA2 Redshift Survey,
which has the wide sky coverage and dense sampling necessary to
delineate voids at redshifts $cz\lesssim10000$~\kms.  These conditions
are not met for the Bo\"otes void, making the definition of Bo\"otes
void galaxies in previous studies harder to interpret.

Using a straightforward density estimation technique, we identified
three large ($\sim\!30$--$50h^{-1}$~Mpc) voids within the
magnitude-limited survey and included all galaxies within these
regions at densities less than the mean ($n<\bar n$).  In addition to
the void galaxies from CfA2, we have also included fainter galaxies in
the same regions from the deeper Century Survey (\cite{cspaper};
hereafter CS) and 15R Survey (\cite{15rcite}).  We thereby gain extra
sensitivity toward the faint end of the void galaxy luminosity
distribution, up to 3 magnitudes fainter than $M_*$.

Covering essentially the entire volume of three distinct voids, our
sample should place improved constraints upon the luminosity, color,
and morphological distributions, and star formation history of void
galaxies.  Moreover, this optically-selected sample should be more
broadly representative than previous void galaxy studies restricted to
emission-line, {\sl IRAS}-selected, and H{\sc i}-selected objects.  We
also conduct a follow-up redshift survey to $m_R=16.13$ in our imaging
survey fields and identify fainter void galaxy companions, akin to the
Szomoru, van Gorkom, \& Gregg (1996) H{\sc i} survey for neighbors of
the Bo\"otes void galaxies.  We thereby probe the small-scale
($\lesssim150h^{-1}$~kpc) environments around galaxies in regions of
large-scale ($5h^{-1}$~Mpc) underdensity.  Here and throughout we
assume a Hubble Constant $H_0 \equiv 100h$~\kms\ Mpc${}^{-1}$.

In \cite{pap1} we introduced the sample and its selection procedure,
described the broadband imaging survey, and examined the variation of
the galaxy luminosity distribution and color distribution with
increasing large-scale underdensity.  The luminosity distribution in
modestly overdense, void periphery regions ($1 < n/\bar n \leq 2$) and
in modestly underdense regions ($0.5 < n/\bar n \leq 1$) are both
consistent with typical redshift survey luminosity functions in $B$ and $R$.
However, galaxies in the lowest-density regions ($n/\bar n \leq 0.5$)
have a significantly steeper LF ($\alpha\sim\!-1.4$).  Similarly, the
\br\ color distribution does not vary with density down to $0.5\bar
n$, but at lower densities the galaxies are significantly bluer.

Here we address the morphology and current star formation (as
indicated by EW(\ha)) of optically-selected galaxies in underdense
regions.  In addition, we describe a deeper redshift survey of the
imaging survey fields designed to reveal nearby companions to the more
luminous void galaxies.  Section \ref{sampsec} reviews the void galaxy
sample selection briefly (cf.~\cite{pap1}) and discusses the selection
of redshift survey targets.  We describe the spectroscopic
observations and data reduction in \S\ref{obsredsec}.  We then analyse
the morphological distribution (\S\ref{mdrressec}) and \ha\ equivalent
width distribution (\S\ref{haressec}) of the sample as a function of
the smoothed large-scale ($5h^{-1}$~Mpc) galaxy number density.
Section \ref{faintressec} describes results from the redshift survey
for close companions.  We conclude in \S\ref{discsec}.
\end{section}
\begin{section}{Sample Selection} \label{sampsec}
\cite{pap1} contains a detailed description of the sample selection
for the imaging and spectroscopic survey, summarized here 
in \S\ref{brightsampsec}.  In \S\ref{faintsampsec} we describe the
selection procedure for a deeper redshift survey of the image survey
fields to identify nearby companions of the sample galaxies.

\begin{subsection}{Imaging and Spectroscopic Survey Sample} \label{brightsampsec}
We use a $5h^{-1}$~Mpc-smoothed density estimator (\cite{gg98}) to identify 
three prominent voids in the CfA2 redshift survey.  We
attempt to include all CfA2 galaxies below the mean density contour
($n < \bar n$) around the voids, as well as fainter galaxies in these
regions from the 15R and Century Surveys.  The apparent magnitude
limit of the CS enables us to include void galaxies with absolute
magnitude $R\lesssim-18$, some three magnitudes fainter than $M_*$.
By restricting our study to galaxies within three of the largest
underdense regions in CfA2 ($\gtrsim 30 h^{-1}$~Mpc diameter), we
minimize the sample contamination by interlopers with large peculiar
velocity.  Table \ref{brighttab} lists the galaxy sample, including 
arcsecond B1950 coordinates, Galactocentric radial velocities, and the
$(n/\bar n)$ corresponding to those locations.
\placetable{brighttab}

We define the galaxies in Table \ref{brighttab} with $(n/\bar n) \leq
1$ as the ``full void sample'', hereafter FVS.  We further examine the
properties of two FVS subsamples: the lowest-density void subsample
(hereafter LDVS) of 46 galaxies with $(n/\bar n) \leq 0.5$, and the
complementary higher-density void subsample (hereafter HDVS) of 104
galaxies with $0.5 < (n/\bar n) \leq 1$.  Our survey also includes
some of the galaxies around the periphery of the voids where $(n/\bar n)
> 1$.  Typically the region surrounding the voids at $1 < n/\bar n
\leq 2$ is narrow, intermediate between the voids and the
higher-density walls and clusters (cf.~\cite{pap1}, Figs.~1--3).
Although our sampling of galaxies in regions with $1 < n/\bar n \leq
2$ is far from complete, we designate these galaxies as a ``void
periphery sample'' (hereafter VPS) to serve as a higher-density
reference for the FVS and its subsamples.  Because the VPS galaxies
are chosen only by their proximity to the voids under study, we should
not have introduced any density-independent selection bias between the
FVS and the VPS.
\end{subsection}
\begin{subsection}{Void Galaxy Field Redshift Survey Sample}\label{faintsampsec}
Most of the volume spanned by the voids of interest has only been
surveyed to the CfA2 magnitude limi, $m_{B}\approx15.5$.
At the 5000--10000~\kms\ distance of these voids, this limiting
magnitude corresponds to an absolute magnitude cutoff of $\sim\!B_*$
or brighter.  To gain information on the presence of fainter
companions to the void galaxies in our study, we use the SExtractor
program (\cite{sexpaper}) to make a list of fainter galaxies on the
$R$-band imaging survey fields.  We define the SExtractor magnitude
$r_{\rm SE}$ as the output MAG\_BEST with ANALYSIS\_THRESH set to
25~mag~arcsec${}^{-2}$ (cf.~\cite{sexpaper}).  We limit the redshift
survey to $r_{\rm SE}=16.1$, the rough limit for efficient redshift
measurement using the FAST spectrograph on the FLWO 1.5~m.  This
magnitude limit is also commensurate with the Century Survey limit
($m_R=16.13$) as well as with the deepest 15R Survey fields in our
study (cf.~\cite{pap1}).

As a check on the reliability of SExtractor magnitudes, we compare
against the isophotal photometry from our imaging survey of the Table
\ref{brighttab} galaxies (\cite{pap1}).  Those $R$-band magnitudes are
determined at the $\mu_B = 26$~mag arcsec${}^{-2}$ isophote; we
denote them $r_{B26}$.  Figure \ref{isosex} shows SExtractor magnitudes
$r_{\rm SE}$ versus $r_{B26}$ for 291 of the 296 galaxies in Table
\ref{brighttab}; the remaining five do not have SExtractor magnitudes
because of saturated $R$-band image pixels (00132+1930, NGC 7311) or
confusion with nearby bright stars (00341+2117, 01193+1531,
23410+1123).  We indicate the linear least-squares fit between
the two magnitude estimates (dotted line), with 11 outliers at
$>2\sigma$ clipped from the fitting.  Because Table \ref{brighttab}
includes fainter 15R and Century Survey galaxies, we have good
calibration down to the $r_{\rm SE}=16.1$ limit of the companion
redshift survey.
\placefigure{isosex}

Figure \ref{isosex} shows that the agreement between $r_{\rm SE}$ and
$r_{B26}$ is excellent over $\approx3.5$~mags.  The scatter about the
fit is only 0.05~mag, comparable to the uncertainty in the $r_{B26}$
magnitudes (\cite{pap1}).  The slope of the fit, $dr_{B26}/dr_{\rm SE}
= 1.043\pm0.004$ indicates that the scale error is negligible.  The
crossover magnitude, for which $r_{\rm SE}=r_{B26}$, is
$15.49\pm0.14$~mag.  The linear fit is sufficiently well-constrained
that our $r_{\rm SE}=16.1$ survey limit corresponds to a limiting
$r_{B26}=16.13\pm0.01$.  This value may be directly compared with
with the similarly-calibrated 15R and Century $r_{B26}$ limits given
in \cite{pap1}.  For the 5000--10000~\kms\ redshift range of the
three voids, this redshift survey therefore includes galaxies brighter
than $R_{B26}\approx -17.4$ to $-18.9$.  Here and throughout the paper
we leave off an implicit ($-5\log h$) when quoting absolute magnitudes.

The $\approx\!11\arcmin$ imaging survey fields are roughly centered on
the target galaxies --- the absolute mean deviation of the pointing
offset is $\approx\!30\arcsec$.  Given this mean offset and the
sample's distribution of angular diameter distance, we estimate that
the mean sky coverage of the redshift survey around the galaxies in
Table \ref{brighttab} drops to 90\% at a projected radius of
$\approx\!115h^{-1}$~kpc .

Table \ref{fainttab} lists the arcsecond B1950 coordinates and
SExtractor magnitudes $r_{\rm SE}$ of the companion redshift survey
targets (sorted by right ascension) as well as the angular separation
of each from its respective ``primary'' (cf.~Tab.~\ref{brighttab}).
Some galaxies in Table \ref{fainttab} have multiple entries because
they neighbor more than one primary.  In some cases the neighbor is
itself a primary from Table \ref{brighttab}; we note this in the
comment field.  There are 211 unique galaxies in Table \ref{fainttab},
which form 250 pairings with primaries from Table \ref{brighttab}.  Of
these 250 pairs, 180 have projected separations $\leq\!115h^{-1}$~kpc.
\placetable{fainttab}
\end{subsection}
\end{section}

\begin{figure}[bp]
{
\plotone{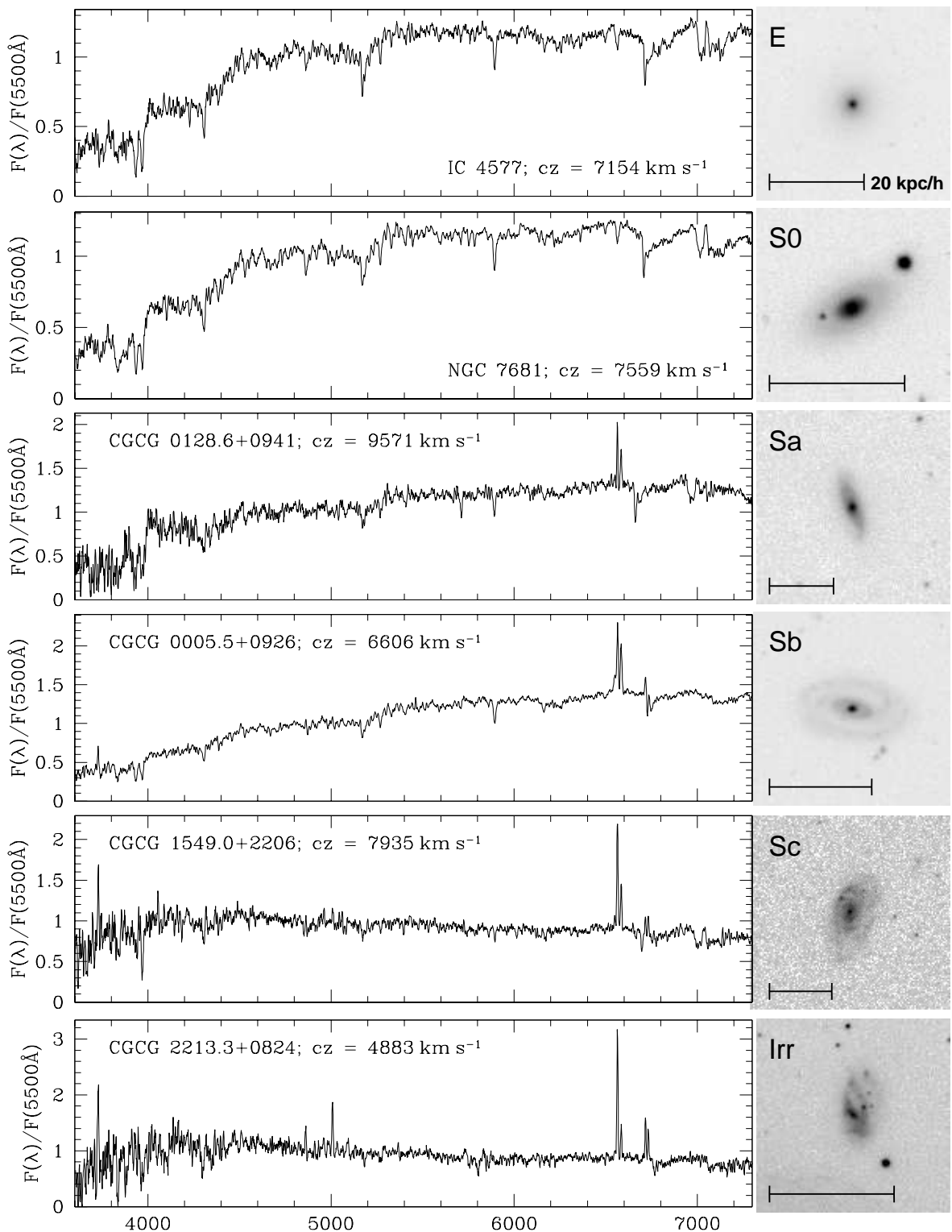}
\figcaption[Grogin.fig2.ps]{
A representative sample of the spectroscopic and
imaging survey data: Longslit spectra and $B$-band images 
for six galaxies spanning early to late morphologies.  The spectra have been
de-redshifted and smoothed to the 6\AA\ FAST spectrograph resolution. 
Galactocentric velocities noted on the spectra
determine the respective $20h^{-1}$~kpc scale bars (for $q_0=0.5$).
\label{imspecfig}}

}
\end{figure}

\begin{section}{Observations and Data Reduction} \label{obsredsec}
\cite{pap1} describes the imaging survey and reductions in detail.
The resulting CCD images from the F. L. Whipple Observatory 1.2~m
reflector have typical exposure times of 300s in $R$ and $2\times300$s
in $B$.  Here we describe 1) a high-S/N spectroscopic survey of the
CGCG and 15R galaxies in the primary sample
(cf.~\S\ref{brightsampsec}); and 2) a deeper redshift survey of
galaxies in the $\sim\!11\arcmin$ $R$-band fields of the imaging
survey (cf.~\S\ref{faintsampsec}).

\begin{subsection}{High-S/N Spectroscopic Survey}\label{brightobssec}
We carried out the spectroscopic survey of CGCG galaxies in our sample
with the FAST longslit CCD spectrograph (\cite{FAST}) on the FLWO 1.5~m
Tillinghast reflector over the period 1995--1998.  We used a
$3\arcsec$ slit and a 300 line mm$^{-1}$ grating, providing spectral
coverage from 3600--7600\AA\ at 6\AA\ resolution.  For the typical
exposure times of 10--20 minutes, we obtained a signal-to-noise ratio
(S/N) in the \ha\ continuum of $\sim\!30$ per 1.5\AA\ pixel.

For the 15R galaxies in our sample, we used the 15R redshift survey
spectra.  These spectra were taken over the period 1994--1996 with
FAST in an identical observing setup to our CGCG spectra.  The
exposure times for these spectra are typically 6--12 minutes, giving
an \ha\ continuum S/N of $\sim\!15$.

The high-S/N spectroscopic survey of CGCG and 15R galaxies is
essentially complete: 100\% for the LDVS; 99\% for the HDVS; and 98\%
for the VPS.  Including the unobserved Century Survey galaxies in the
accounting, the overall spectroscopic completeness is 98\% for the
LDVS, 95\% for the HDVS, and 95\% for the VPS.

All spectra were reduced and wavelength-calibrated using standard IRAF
tasks as part of the CfA spectroscopic data pipeline (cf.~\cite{km98}).
We flux-calibrate the resulting 1-D spectra with spectrophotometric
standards (\cite{mas1,mas2}) taken on the same nights.  Because these
spectra were observed as part of the FAST batch queue, the observing
conditions were not always photometric.  We therefore treat the flux
calibrations as relative rather than absolute, and only quote
equivalent widths rather than line fluxes.
 
We next de-redshift each spectrum using the error-weighted mean of
cross-correlation radial velocities found with FAST-customized
emission- and absorption-line templates (cf.\@ \cite{km98,gg98}).  Our
redshifts from the high-S/N CGCG spectra supersede the previous CfA
redshift survey values and are reflected in the recent Updated Zwicky
Catalog (hereafter UZC: \cite{uzc}).  We do not correct the spectra
for reddening (intrinsic or Galactic), but note that the majority of
sample galaxies are at high Galactic latitudes where Galactic
reddening is minimal.  Figure \ref{imspecfig} shows a representative
subset of our reduced imaging and spectroscopic data: $B$-band images
and corresponding spectra for a range of early- to late-type galaxies.
\placefigure{imspecfig}

We make a first pass through the de-redshifted spectra with SPLOT,
fitting blended Gaussian line profiles to the \ha\ and N{\sc ii}
lines, and also fitting H$\gamma$ and H$\delta$ for later
Balmer-absorption correction.  We note that the FAST spectral
resolution of 6\AA\ allows clean deblending of \ha\ $\lambda$6563\AA\
from the adjacent [N{\sc ii}] $\lambda\lambda$6548,6584\AA\ lines.
Using the first-pass line centers and widths, we make a second pass
through the spectra to determine the equivalent widths and associated
errors via direct line- and continuum-integration rather than
profile-fitting.

We apply an approximate correction to EW(\ha) for Balmer absorption by
using the greater of EW(H$\gamma$) and EW(H$\delta$) if detected in
absorption at $\geq1\sigma$.  We note that only one of the 277
galaxies in our spectroscopic sample has Balmer absorption exceeding
5\AA\ (CGCG 0109.0+0104), and this object has strong emission lines.
There appear to be no ``E+A'' galaxies in our sample, which is not
surprising given the small fraction ($\approx0.2$\%) of such objects
in the local universe (\cite{zab96}).  Table \ref{brighttab} includes
the resulting \ha\ equivalent widths and their errors.
\end{subsection}
\begin{subsection}{Redshift Survey of Void Galaxy Fields} \label{faintobssec}
We observed the redshift survey targets of Table \ref{fainttab} with the
FAST longslit CCD spectrograph (\cite{FAST}) on the FLWO 1.5~m
Tillinghast reflector over the period June 1996--November 1997 as part
of the FAST batch queue.  The exposure times ranged from 5--20
minutes, with a median of 12 minutes.  The observing setup, as well as
the spectrum reduction, wavelength-calibration, and redshift
extraction, were identical to the high-S/N spectroscopic survey
(\S\ref{brightobssec}).

Of the 211 galaxies in Table \ref{fainttab}, we include new
measurements for 83.  Another 46 are members of the primary sample,
and thus have known redshifts.  For the remainder, we obtained 35
redshifts from the 15R and Century Surveys, the UZC (\cite{uzc}), ZCAT
(\cite{zcat}), and NED\footnote{The NASA/IPAC Extragalactic Database
(NED) is operated by the Jet Propulsion Laboratory, California
Institute of Technology, under contract with the National Aeronautics
and Space Administration.}.  The median uncertainty for the
redshifts presented in Table \ref{fainttab} is 21~\kms.

We lack a redshift for 47 galaxies, a completeness of 78\%.  These
galaxies are bunched near the magnitude limit --- 36 of the 47 have
$r_{\rm SE} > 15.5$.  The survey completeness by field is somewhat
greater because many of the imaging survey fields have no follow-up
targets: 89\% of the LDVS fields are fully surveyed, 85\% of the HDVS,
and 88\% of the VPS.

Of the 165 galaxies in Table \ref{fainttab} which have a projected
radius $D_p \leq 115h^{-1}$~kpc from a Table \ref{brighttab} galaxy,
we have redshifts for 129 --- again 78\% complete.  The completeness
by field is slightly larger when restricted to $D_p<115h^{-1}$~kpc
because more of the fields have no targets: 91\% of the LDVS fields
are fully surveyed for $D_p\leq115h^{-1}$~kpc, 88\% of the HDVS, and
90\% of the VPS.
\end{subsection}
\end{section}

\begin{figure}[bp]
{
\plotone{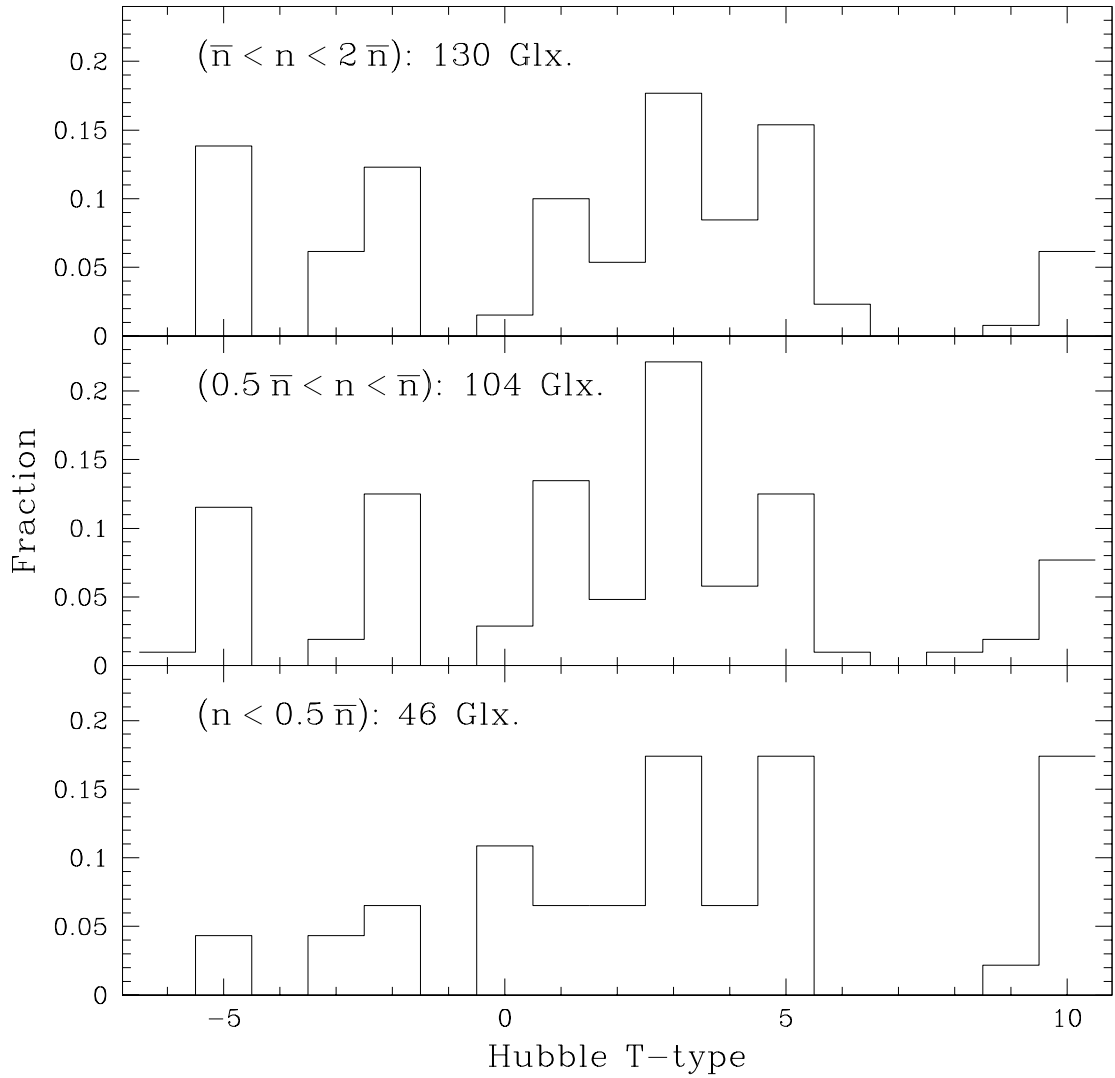}
\caption{
Histograms of the revised morphological type
$T$ for the VPS (upper panel), the HDVS (middle panel), and the LDVS
(lower panel).
\label{tridentt}}
}
\end{figure}

\begin{figure}[bp]
{
\plotone{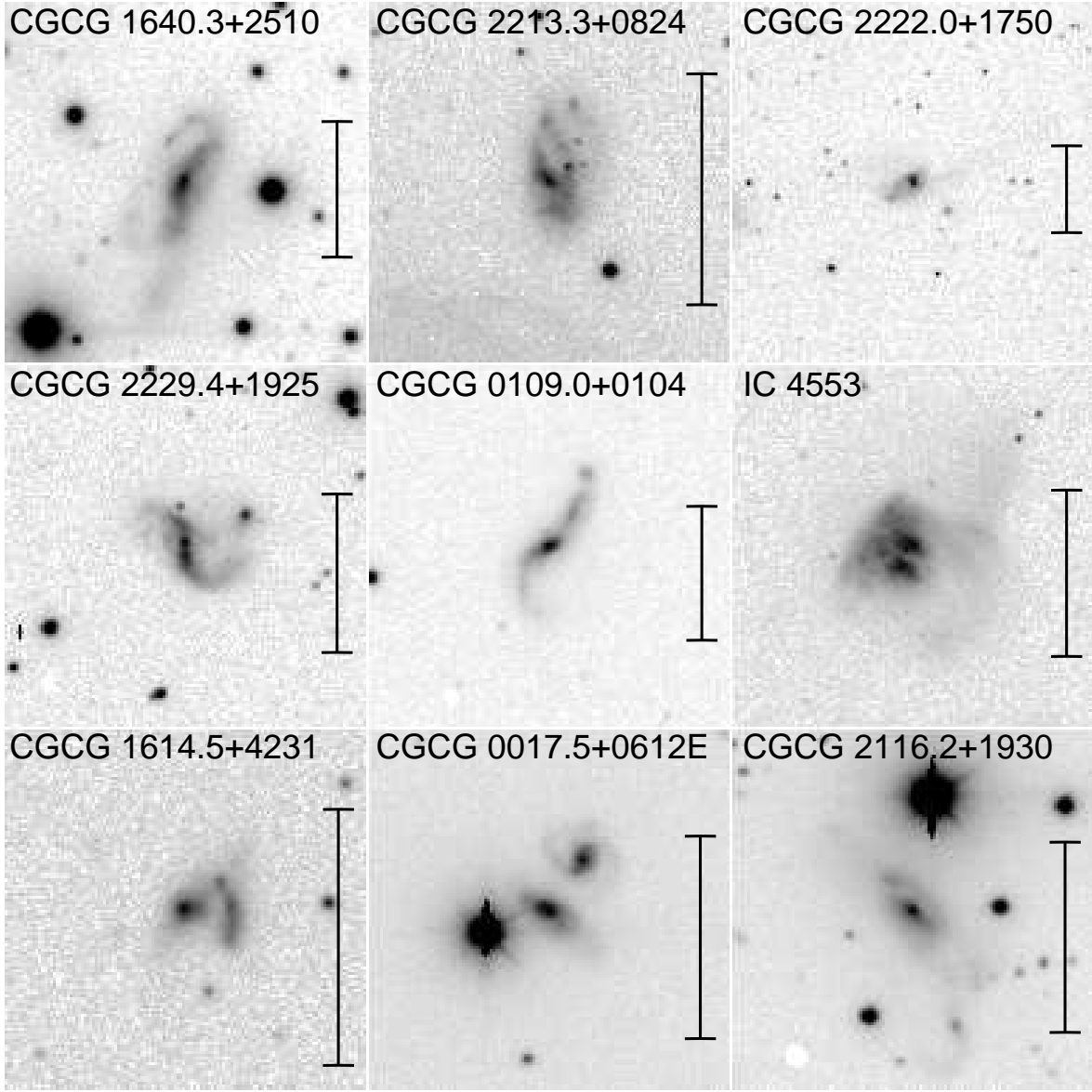}
\caption{
A mosaic of 9 $B$-band images of galaxies in the
LDVS ($n \leq 0.5 \bar n$) which show disturbed morphology and/or
evidence of interaction.  The scale bar in each image represents
$20h^{-1}$~kpc at the galaxy redshift.
\label{ldinter}}
}
\end{figure}

\begin{figure}[bp]
{
\plotone{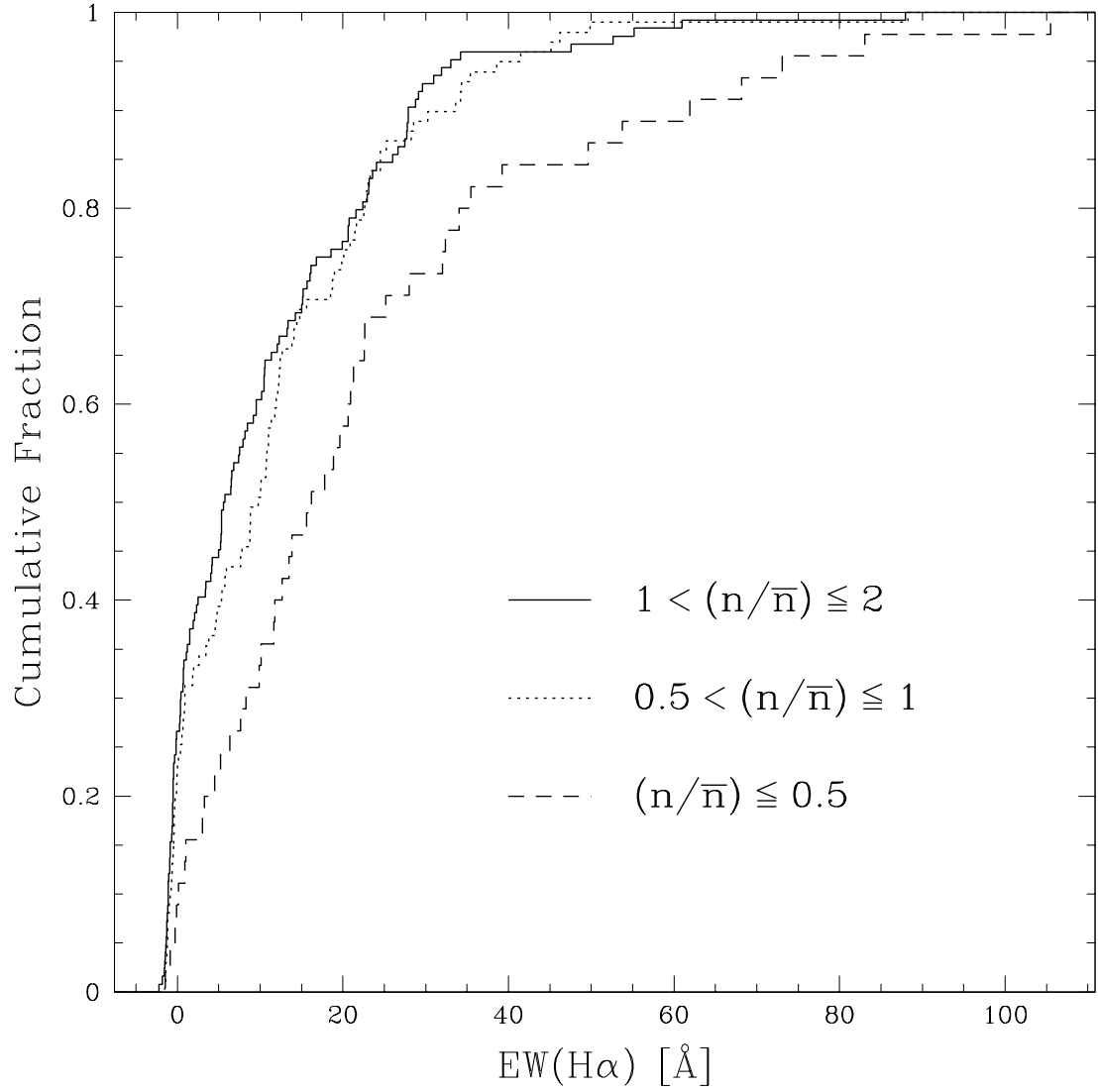}
\caption{
\ha\ equivalent width versus density $(n/\bar n)$.
\label{tridenha}}

}
\end{figure}

\begin{figure}[bp]
{
\plotone{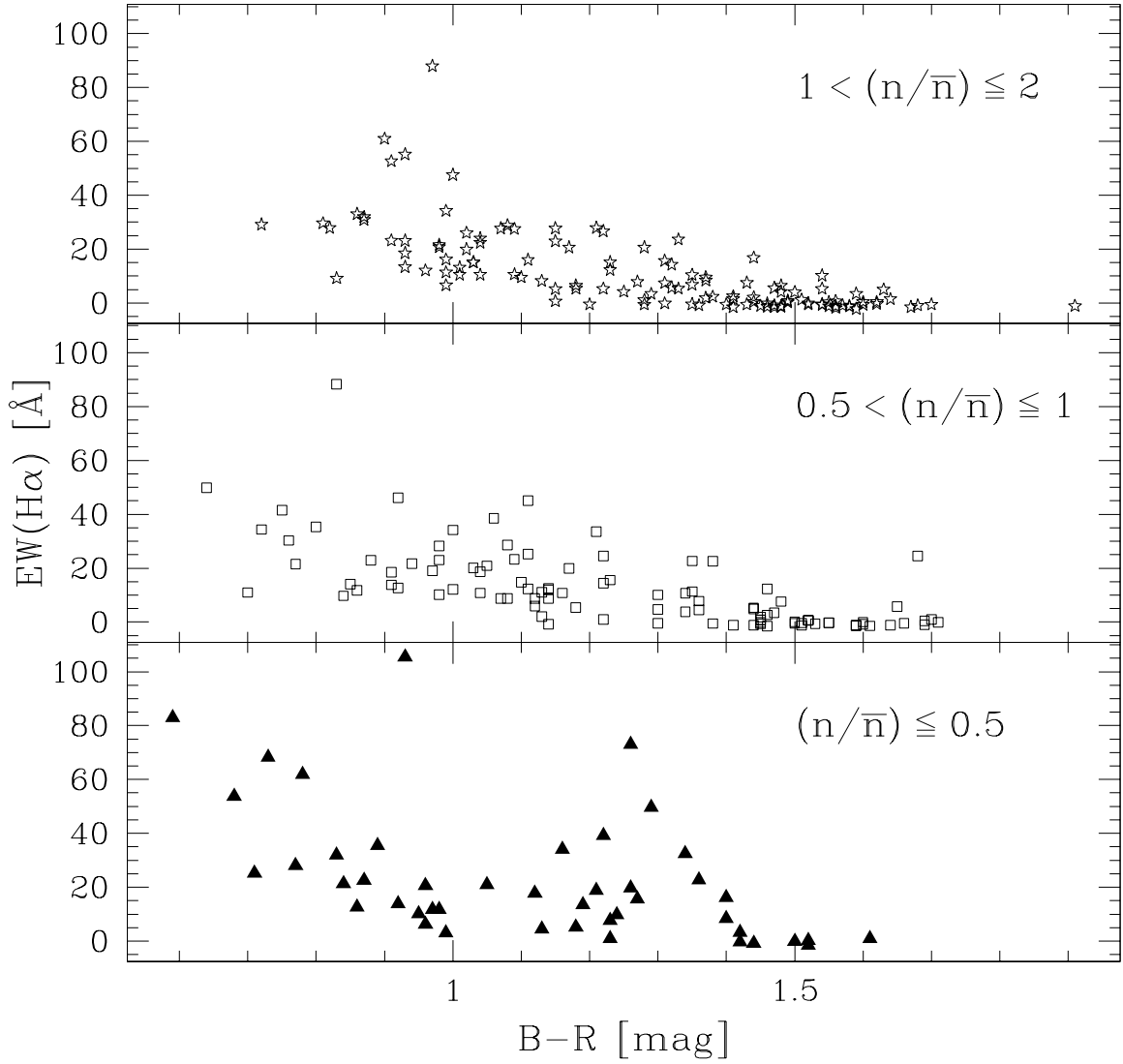}
\caption{
\ha\ equivalent width versus absolute \br\ color
for the VPS (top), the HDVS (middle), and the 
LDVS (bottom).
\label{tridencolha}}

}
\end{figure}

\begin{figure}[bp]
{
\plotone{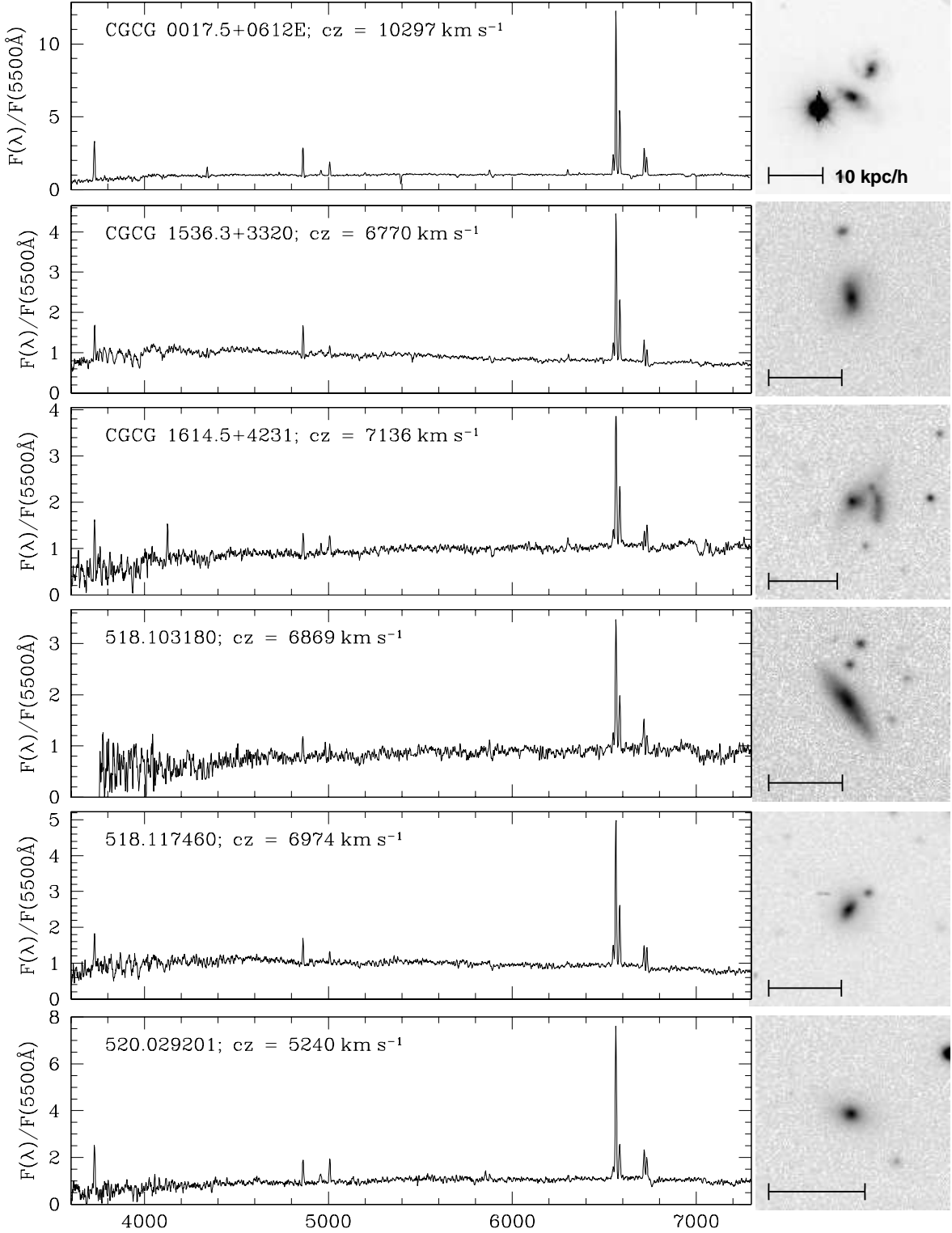}
\caption{ 
Longslit spectra and $B$-band images for the six LDVS galaxies
with red colors ($\br>1.2$) and EW(\ha)$ > 20$\AA.  The spectra have
been de-redshifted and smoothed to the 6\AA\ FAST spectrograph
resolution.  Galactocentric velocities noted on the spectra
determine the respective $10h^{-1}$~kpc scale bars (for $q_0=0.5$).
\label{redldbigha}}
}
\end{figure}

\begin{figure}[bp]
{
\plotone{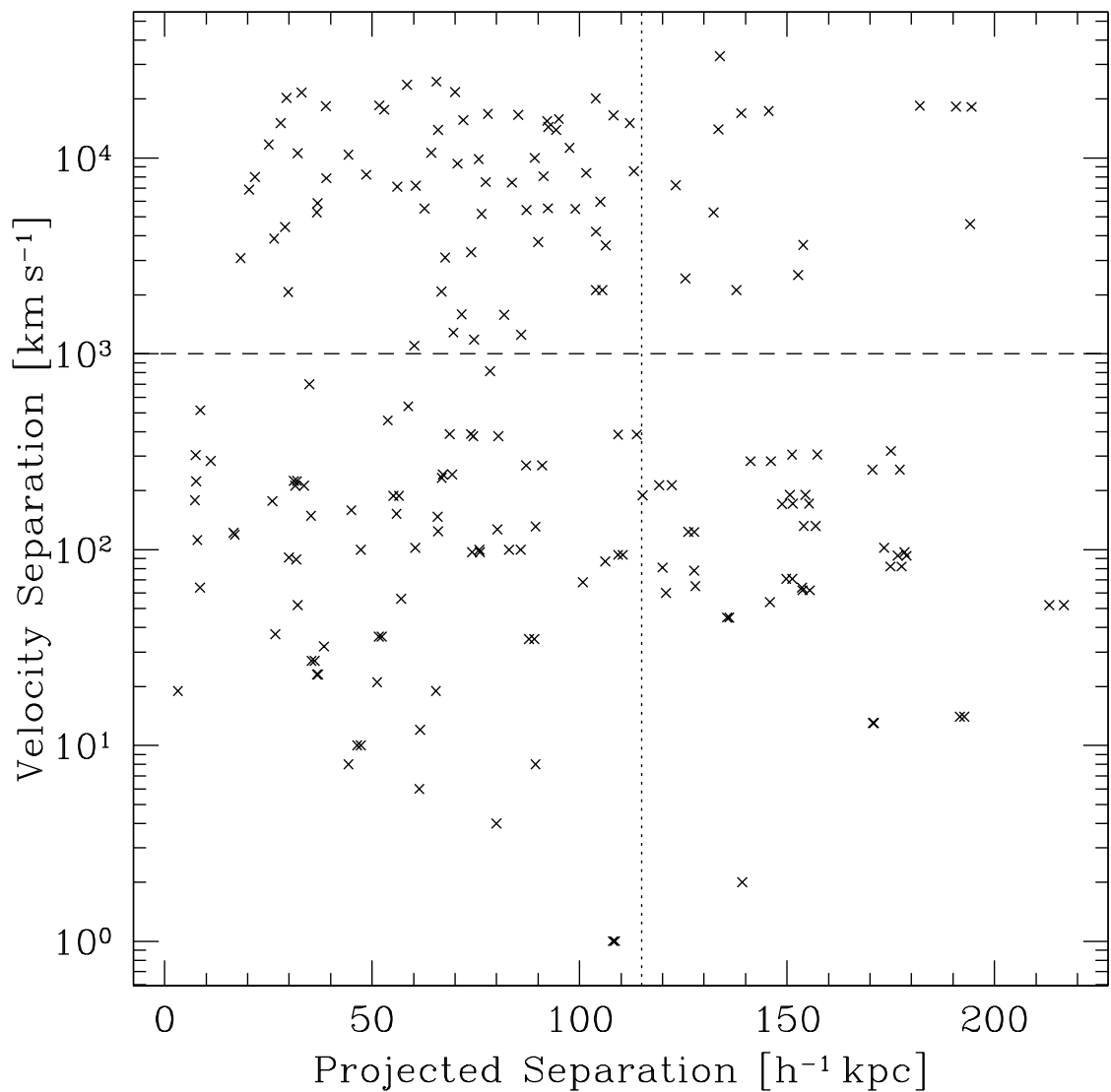}
\caption{
Projected separations (in
$h^{-1}$~kpc) and absolute velocity separations for 198 pairs
(Table \ref{fainttab}).  The redshift survey sky coverage is 
$\approx\!90$\% to a projected radius of $115h^{-1}$~kpc (dotted
line).  We only consider a pair to be associated if the velocity
separation is $<1000$~\kms\ (dashed line). 
\label{pvsepfig}}
}
\end{figure}

\begin{figure}[bp]
{
\plotone{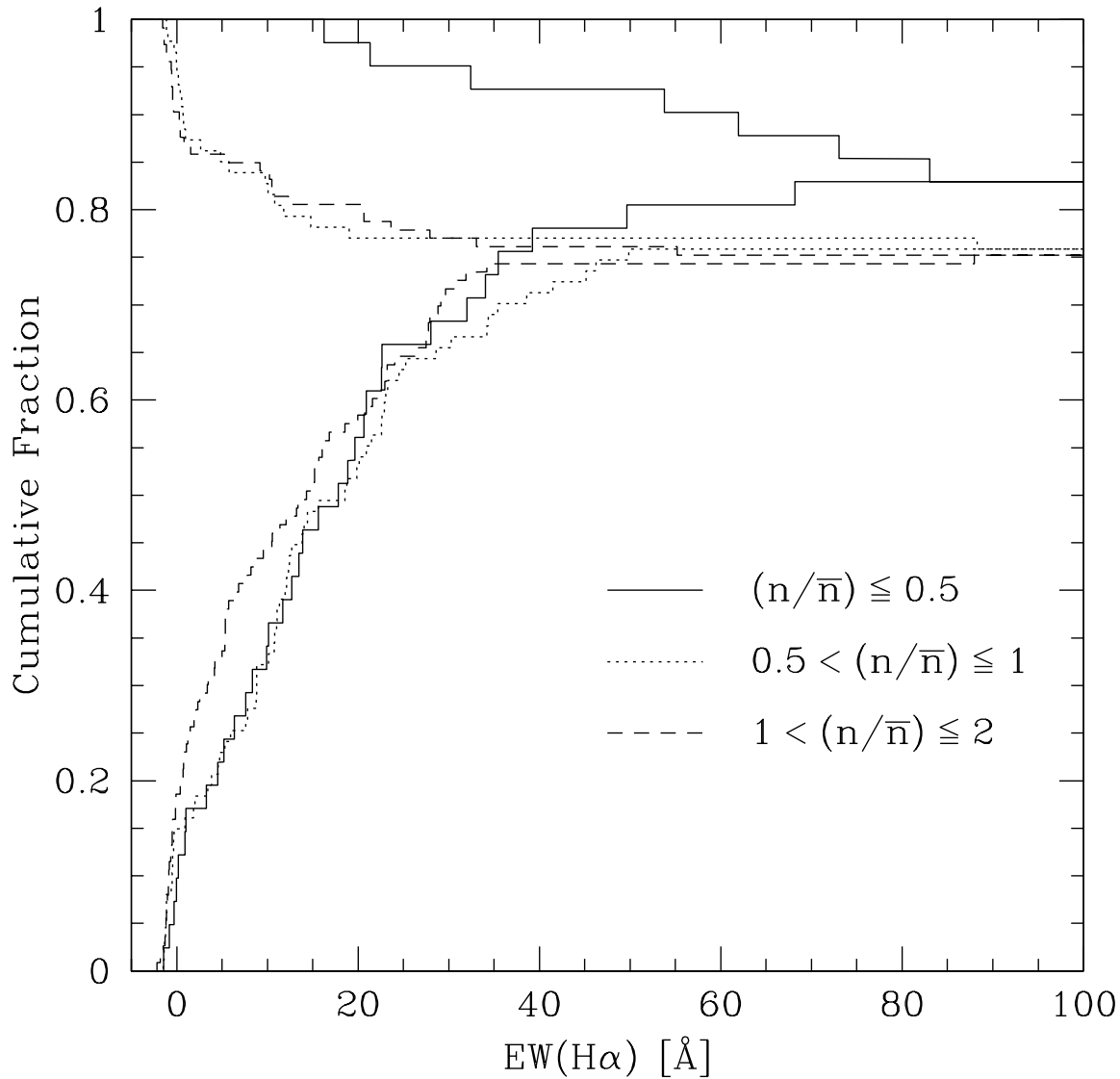}
\caption{ 
Cumulative distribution function of
\ha\ equivalent widths segregated by density subsample (VPS: dashed;
HDVS: dotted; LDVS: solid) and by presence of a
nearby companion.  Unpaired galaxies, having no $m_R<16.13$
companions within 1000~\kms\ and $115h^{-1}$~kpc, comprise the 
lower curves; galaxies with at least one such companion
comprise the upper curves.  Upper and lower curves converge at the
fraction of unpaired systems with measured EW(\ha).
\label{dualcdf}}
}
\end{figure}

\begin{figure}[bp]
{
\plotone{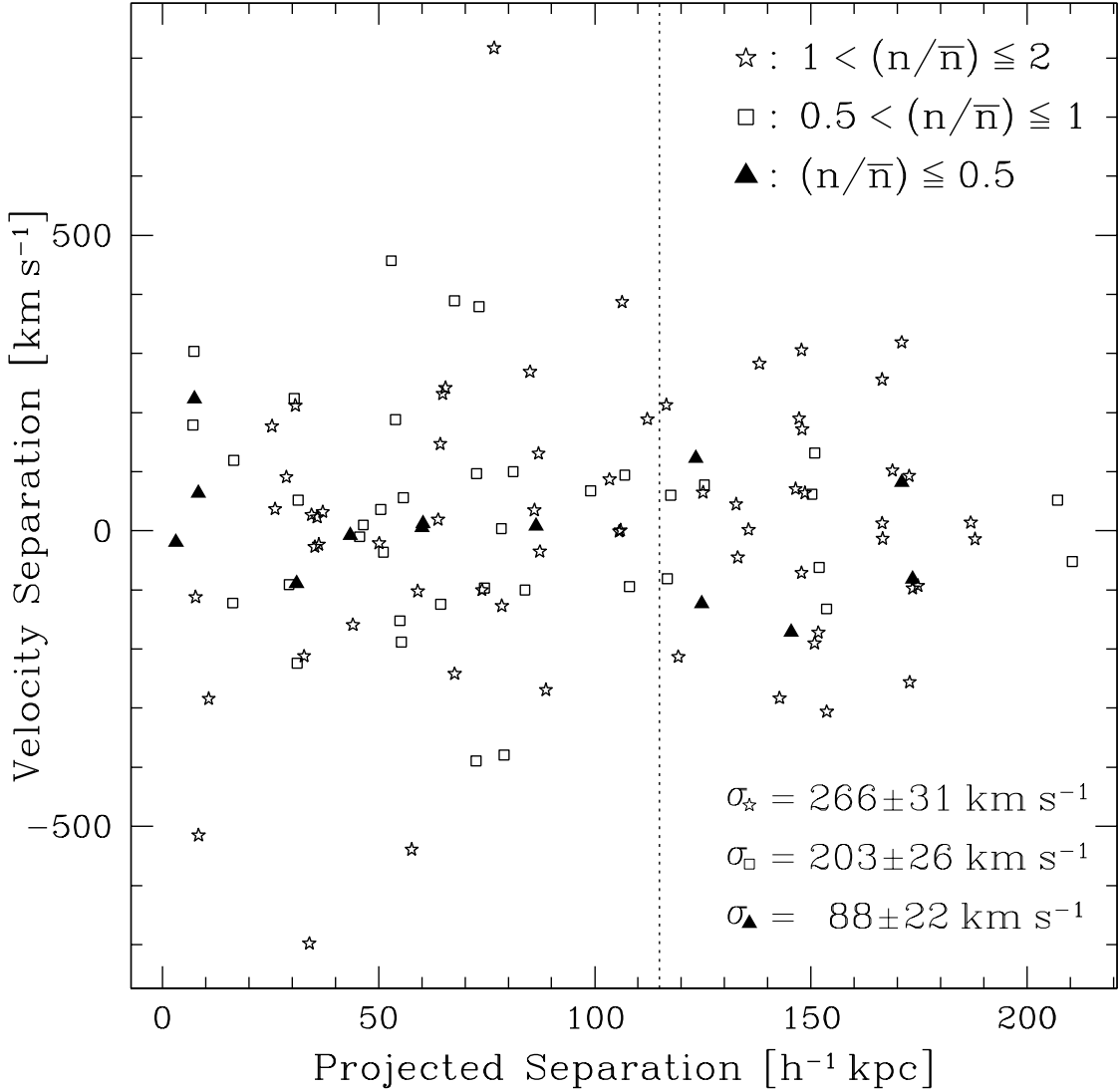}
\caption{ 
Radial velocity separation as a function of
projected separation for galaxies in Table \ref{fainttab} with
$|\Delta cz| < 1000$~\kms.  The symbols for each density subsample 
are at upper right.  The redshift survey sky coverage is 
$\approx\!90$\% to a projected radius of $115h^{-1}$~kpc (dotted
line).  The variances at lower right are for pairs within this radius.
\label{velsepdenfig}}}
\end{figure}

\begin{section}{Morphology-Density Relation} \label{mdrressec}
One of us (N.A.G.)  classified the morphologies of the entire sample
by eye from the $B$ CCD images.  The median
seeing during the observations was $\approx2\farcs0$ and varied
between $1\farcs4$ and $3\farcs3$.  The target galaxies, all with
$5000 \lesssim cz \lesssim 10000$~\kms, are typically
$\lesssim90\arcsec$ in diameter and are roughly centered within the
$\approx11\arcmin$ CCD fields ($\approx0\farcs65$ per
$2\times2$-binned pixel).  We assign each galaxy in Table
\ref{brighttab} a ``$T$-type'' from the revised morphological system,
with the caveat that we list both irregular and peculiar galaxies as
$T=10$. From repeatability of classification for galaxies imaged on
multiple nights, as well as from independent verification of the
classifications by several ``experts'' (J. Huchra, M. Kurtz,
R. Olowin, and G. Wegner), we estimate that the classifications are
accurate to $\sigma_T\sim\pm1$ for the CGCG galaxies and
$\sigma_T\lesssim\pm2$ for the fainter (and typically smaller in
angular size) 15R and Century galaxies.

We plot (Fig.~\ref{tridentt}) the histograms of revised morphological
type $T$ for the VPS (top), the HDVS (middle), and the LDVS (bottom).
The VPS and the HDVS are very similar in their
morphological mix, with an early-type fraction 
of $\approx30$\% (Tab.~\ref{morphtab}).  A chi-square
test of these two histograms yields a $78\%$ probability of the null
hypothesis that the VPS and HDVS have a consistent underlying
morphological distribution.  \placefigure{tridentt} \placetable{morphtab}

In contrast, Figure \ref{tridentt} shows that the morphological mix
changes significantly at the lowest densities (LDVS).  There is a
notable increase in the fraction of Irr/Pec galaxies and a
corresponding decrease in the early-type fraction
(Tab.~\ref{morphtab}).  A chi-square test between the VPS and LDVS
morphology histograms gives only a 7\% probability that these two
samples reflect the same underlying morphological mix.  These two
samples are well-separated in surrounding density --- the uncertainty
in the $5h^{-1}$~Mpc density estimator is $\lesssim0.1$ at the
distance of the three voids in this study (\cite{gg98}).  Clearly 
a larger sample is desirable to better establish the morphological 
similarity between the VPS and HDVS, and their morphological contrast
with the LDVS.

The incidence of qualitatively disturbed or interacting systems
appears somewhat larger for the LDVS: $\sim\!35\pm10\%$, compared with
$\sim\!20\pm5\%$ for the galaxies at larger $n$.  We show
(Fig.~\ref{ldinter}) a mosaic of 9 $B$-band images of LDVS galaxies
which are probable interactions.  Notable among these interacting void
galaxies is the spectacular object IC 4553 (Arp 220), the prototype
(and nearest) ultraluminous IR galaxy.  An increase in disturbed
galaxies at the lowest global densities seems counterintuitive.  We
show (\S\ref{faintressec}) that the effect may result from a low
small-scale velocity dispersion in these regions.
\placefigure{ldinter}
\end{section}
\begin{section}{EW(\ha)-Density Relation} \label{haressec}
Figure \ref{tridenha} shows the cumulative distribution function (CDF)
of \ha\ equivalent width for the three different density regimes: the
VPS (dashed), the HDVS (dotted), and the LDVS (solid).  The similarity
between the VPS and HDVS is evident, with a Kolmogorov-Smirnov (K-S)
probability of 32\% that the galaxies in these two density regimes
have a consistent underlying distribution of EW(\ha).  Given the
similar fraction of early-type galaxies in these two samples
(Fig.~\ref{tridentt}), it is not surprising that we see a similar
fraction of absorption-line systems ($\approx35\%$).  If this
absorption-line fraction is representative of the overall survey at
similar densities, then void galaxy studies drawn from emission-line
surveys miss roughly one-third of the luminous galaxies in regions of
modest global underdensitiy.  Figure \ref{tridenha} shows that there
are galaxies even at $n\leq0.5\bar n$ with old stellar populations and
no appreciable current star formation.  \placefigure{tridenha}

The shift toward late-type morphology in the LDVS
(Fig.~\ref{tridentt}) is mirrored by a shift toward larger \ha\
eqivalent widths (Fig.~\ref{tridenha}).  Absorption-line systems
are less than half as abundant at $n\leq0.5\bar n$ ($\approx15$\% of
the total); strong ELGs with EW(\ha) $> 40$\AA\ are more than
three times as abundant.  The K-S probability of the LDVS and VPS
representing the same underlying distribution of EW(\ha) is only
0.4\%.  The probability rises to 3\% between the LDVS and HDVS.

Figure \ref{tridencolha} shows EW(\ha) as a function of the galaxies'
\br\ colors --- the shift toward bluer galaxies in the LDVS is clear
(cf.~\cite{pap1}).  The red galaxies are predominantly absorption-line
systems, with the notable exception of several galaxies in the LDVS
with $\br \gtrsim 1.2$ and EW(\ha) $\gtrsim 20$\AA.  Figure
\ref{redldbigha} displays these galaxies' spectra and $B$-band images.
Only two have bright nearby companions (CGCG 0017.5+0612E, CGCG
1614.5+4231), but the others have possible faint companions.  All
appear to be disk systems, and the red colors probably result from
internal reddening by dust; Balmer decrements are in the range 7--9
for these galaxies compared with the typical value of $\approx\!2.8$
for case-B recombination.  \placefigure{tridencolha}
\placefigure{redldbigha}
\end{section}
\begin{section}{Void Galaxy Companions} \label{faintressec}
Here we discuss various results stemming from our deeper redshift
survey of the imaging survey fields to $m_R = 16.13$
(cf.~\S\ref{faintobssec}).  We determine the incidence of close
companions as a function of density environment, examine
the relationship between the presence of companions and the
distribution of EW(\ha) versus density, and measure the velocity
separation of the close companions as a function of $(n/\bar n)$.

\begin{subsection}{Incidence of Close Companions and Effect on EW(\ha)}
Figure \ref{pvsepfig} shows the projected separations (in
$h^{-1}$~kpc) and absolute velocity separations for all entries in
Table \ref{fainttab} with measured redshift.  A galaxy in Table
\ref{fainttab} counts as a companion if the velocity separation from
the primary is $<1000$~\kms\ (dashed horizontal line).  This velocity
cutoff is generous, but the gap in Figure \ref{pvsepfig} at $|\Delta
cz| \sim 500$--2000~\kms\ leads us to expect few interlopers.  Because
the sky coverage of the neighbor redshift survey becomes increasingly
sparse at projected separations $D_p \gtrsim 115h^{-1}$~kpc (dotted
vertical line), we repeat the analyses in this section with and
without the added companion criterion $D_p \leq 115h^{-1}$~kpc.
\placefigure{pvsepfig}

Table \ref{comptab} lists the fraction of galaxies in the various
density groupings which are classified as ``unpaired'' (zero
companions as defined above) and ``paired'' (at least one such
companion), or cannot yet be classified because of one or more missing
redshifts in Table \ref{fainttab}.  The fraction of paired galaxies
decreases consistently across all density subsamples by $\sim\!25$\%
under the restriction $D_p \leq 115h^{-1}$~kpc .  Table \ref{comptab}
shows that the fraction of paired galaxies is largely insensitive to
the global density environment.  Szomoru et al.~(1996), who detected
29 companions around 12 Bo\"otes void galaxies in H{\sc i}, also noted
this tendency of void galaxies to be no less isolated on these small
scales than galaxies at higher density.  
\placetable{comptab}

We investigate the relationship between close companions and recent
star formation in void galaxies by comparing EW(\ha) CDFs for paired
versus unpaired galaxies.  Figure \ref{dualcdf} shows a dual-CDF plot
of the paired and unpaired galaxies' EW(\ha) (with the companion
restriction $D_p \leq 115h^{-1}$~kpc): the unpaired galaxies' CDF
increases from the bottom; the paired galaxies' CDF decreases from the
top.  As in Figure \ref{tridenha}, we distinguish between the VPS
(dashed), HDVS (dotted), and LDVS (solid).  The overall fraction of
galaxies exceeding a given EW(\ha) is now represented by the interval
{\sl between} the upper and lower curves.  For example, the excess of
high-EW(\ha) galaxies in the LDVS is reflected here in the slower
convergence of upper and lower solid curves until large EW(\ha).  The
upper and lower curves converge to the fraction of galaxies with
EW(\ha) measurements but without detected companions.  As the overall
unpaired fraction is similar for each density subsample
(cf.~Tab.~\ref{comptab}), it is reassuring that the three sets of
curves converge at similar levels.  \placefigure{dualcdf}

Inspection of the lower curves of Figure \ref{dualcdf} reveals that
the unpaired galaxies have much the same distribution of EW(\ha),
regardless of the global density environment.  The K-S probabilities
(Tab.~\ref{ksisotab}) confirm this impression.  In contrast, the
galaxies with companions have a very different EW(\ha) distribution
depending on their density environment.  Absorption-line systems
account for $\approx\!20$\% of the paired galaxies in both the VPS and
the HDVS, while there are essentially no LDVS absorption-line systems
with companions.  A K-S comparison of the EW(\ha) distribution of the
paired-galaxy LDVS with either the HDVS or VPS reveals a highly
significant ($\sim\!4\sigma$) mismatch, whereas the EW(\ha)
distributions for the paired-galaxy HDVS and VPS are consistent
(Tab.~\ref{ksisotab}).  \placetable{ksisotab}
\end{subsection}
\begin{subsection}{Pair Velocity Separation vs.~Density} \label{pairvelsec}
Figure \ref{velsepdenfig} shows the radial velocity separation $\Delta
cz$ versus projected separation for entries in Table \ref{fainttab}
with $|\Delta cz| < 1000$~\kms.  Clearly the velocity separations of
the LDVS pairs (solid triangles) are much smaller than either the HDVS
pairs (open squares) or the VPS pairs (open stars).  For the pairs
which fall within the $\approx\!90$\% coverage radius of $115h^{-1}$~kpc
(dotted line), the dispersion in velocity separation $\sigma_{\Delta
cz}$ is $88\pm22$~\kms\ for the LDVS, compared with $203\pm26$~\kms\
for the HDVS and $266\pm31$~\kms\ for the VPS (Table \ref{comptab}).
These values for $\sigma_{\Delta cz}$ do not vary by more than the
errors if we include the points in Figure \ref{velsepdenfig} with projected 
separations $>115h^{-1}$~kpc (``All $D_p$'' in Tab.~\ref{comptab}).
\placefigure{velsepdenfig} 

An F-test (Tab.~\ref{ftesttab}) between the respective $\Delta cz$
distributions for companions with $D_p \leq 115h^{-1}$~kpc gives a low
probability $P_{\rm F}$ that the LDVS could have the same underlying
$\Delta cz$ variance as the HDVS ($P_{\rm F} = 3.2\%$) or the VPS
($P_{\rm F} = 0.59\%$).  The difference between HDVS and VPS
dispersions is not significant at the $2\sigma$ level ($P_{\rm F} =
13\%$).  The F-test probabilities using all $D_p$ from Table
\ref{fainttab} are almost identical (Tab.~\ref{ftesttab}).  We discuss
the implications of the velocity dispersion variation in
\S\ref{discsec}.  \placetable{ftesttab}
\end{subsection}
\end{section}
\begin{section}{Summary and Conclusions} \label{discsec}
Our $B$- and $R$-band CCD imaging survey and high-S/N longslit
spectroscopic survey of $\sim\!300$ galaxies in and around three
prominent nearby voids have enabled us to examine the morphologies and
star-formation history (in terms of EW(\ha)) as a function of the
global density environment for $n\leq2\bar n$.  These studies
complement our earlier examination of the luminosity and \br\
color distributions of the same galaxies (\cite{pap1}).  We have also
described an additional redshift survey of projected ``companions'' to
$m_R=16.13$ which probes the very local environments
($\lesssim\!150h^{-1}$~kpc) around these galaxies within globally
($5h^{-1}$~Mpc) low-density regions.

Our analysis of the CCD $B$ morphologies and \ha\ linewidths reveals:
\begin{enumerate}
\item The distribution of galaxy morphologies varies little with 
large-scale ($5h^{-1}$~Mpc) density environment over the range 
$0.5<(n/\bar n)\leq2$, with a consistent fraction of early types 
($\approx\!35$\% with $T<0$).  The distribution of
\ha\ equivalent widths, indicative of star-formation history, is
similarly invariant with large-scale density over this range. 
\item At large-scale densities below half the mean, both the
morphology and EW(\ha) distributions deviate at the 2--3$\sigma$ level
from the higher-density subsamples.  There is a reduction in the
early-type fraction (down to $\approx15$\%) and a corresponding
increase in the fraction of irregular/peculiar morphologies.  More of
these galaxies show active star formation; even several of the redder
galaxies ($\br > 1.2$) have EW(\ha)$ > 20$\AA.  Many of the void
galaxies, particularly at the lowest densities, show evidence of
recent or current mergers/interaction.
\end{enumerate}

The results here and in \cite{pap1} can be combined into a consistent
picture of the trends in galaxy properties with large-scale
underdensity.  In \cite{pap1} we showed that the luminosity function
in the voids at $\gtrsim\!0.5\bar n$ is consistent with typical
redshift survey LFs; at densities below $0.5\bar n$ the LF faint-end
slope steepens to $\alpha \sim -1.4$.  Recent studies point to a
type-dependent galaxy LF with steeper faint-end slope for the late
morphologies in CfA2+SSRS2 (\cite{marz98}) and for the ELGs in the Las
Campanas Redshift Survey (\cite{lcrs2}).  We might therefore expect our
morphological and spectroscopic distributions to vary little over the
range $0.5<(n/\bar n)\leq 2$ and to shift toward late types and ELGs
at the lowest densities.  This trend is exactly what we observe
(cf.~\S\ref{mdrressec})

Furthermore, we found in \cite{pap1} that the \br\ color distribution of
our sample at densities $\gtrsim0.5\bar n$ is consistent with the
overall survey and shifts significant toward the blue for $n \leq 0.5\bar n$.
This trend is consistent with the observed shift at $(n \leq 0.5\bar n)$ toward 
late-type and high-EW(\ha) galaxies, which are typically bluer than
early-type and absorption-line systems.

To ascertain whether these changes in galaxy properties below $(n/\bar
n)\approx0.5$ are caused by variations in the {\sl local}
redshift-space environments of these galaxies (within
$\lesssim\!150h^{-1}$~kpc), we have carried out a deeper redshift
survey of the imaging survey fields, to $m_R = 16.13$.  The relative
fractions of unpaired versus paired galaxies (paired galaxies are
closer than $115h^{-1}$~kpc projected and 1000~\kms~in redshift) does
not significantly vary with the degree of global underdensity.
Furthermore, the distribution of EW(\ha) for the unpaired galaxies
varies little between our density subsamples.  However, the galaxies
at the lowest densities ($n/\bar n \leq 0.5$) which have companions
are invariably ELGs.  At higher global densities, roughly
$\approx\!20\%$ of the galaxies with companions are absorption-line
systems.  This difference in paired galaxy EW(\ha) with density is
significant at the $4\sigma$ level.

Our companion redshift survey further reveals that the pair velocity
separation decreases significantly ($3\sigma$) at the lowest
densities, in support of theoretical predictions (e.g., \cite{nar98})
that within the voids the velocity dispersion among galaxies should
decline.  Because
associated galaxies at smaller velocity separations should have more
effective interactions, the excess strong ELGs and disturbed
morphologies at {\sl global} underdensities $(n \leq 0.5\bar n)$ may
be ascribed to {\sl local} influences.

These results argue for a hierarchical galaxy formation scenario where
the luminous galaxies in higher-density regions formed earlier than at
much lower density (e.g., \cite{kauff96}).  The older galaxies at
higher-density would typically have less gas and dust at the present
epoch, and thus show less active star formation even in the presence
of nearby companions.  In the voids, where the luminous galaxies
presumably formed more recently, there should be more gas and dust
present for active star formation triggered by nearby companions.  As
the EW(\ha) distributions are almost identical for the {\sl unpaired}
galaxies at different global density, we conclude that the {\sl local}
environment, i.e.\@ the presence or absence of nearby
($\lesssim\!150h^{-1}$~kpc) companions, has more influence upon the
current rate of star formation in these regions.  In a future paper we
hope to clarify the relationship between global underdensity and
galaxy age (and metallicity) by studying the absorption-line indices
of our high-S/N spectra (cf.~\cite{trag98} and references therein).
We may thereby use low-density regions to test the prediction by
Balland, Silk, \& Schaeffer (1998) that non-cluster ellipticals must
have all formed at high redshift ($z\gtrsim2.5$).

Although the sample of void galaxies described here is much larger
than previous studies of at most a few dozen objects
(e.g.~\cite{cwh97,sz96a,thor95,w95}), the distinctions in galaxy
properties we observe at the lowest densities are based upon $<50$
objects.  We should like to increase the LDVS sample size to improve
our statistics.  One avenue is to include all other low-density
portions of CfA2 and the comparably deep SSRS2 survey (\cite{ssrs2}).
Unfortunately the centers of voids are very empty, at least to
$m_B\approx15.5$, and we would only expect to increase the low density
sample thereby to $\sim\!100$ galaxies.  Because the global density
estimator requires a well-sampled, contiguous volume with dimensions
$\gg5h^{-1}$~Mpc, growing the LDVS sample significantly is contigent
upon deeper, wide-angle redshift surveys like the Sloan Digital Sky
Survey (\cite{sdss}) and the 2dF Galaxy Redshift Survey (\cite{2df}).

\acknowledgments We acknowledge the FAST
queue observers, especially P. Berlind and J. Peters, for their help
in obtaining our spectra; E. Barton, A. Mahdavi, and S. Tokarz for
assistance with the spectroscopic reduction; and D. Fabricant for the
FAST spectrograph.  We give thanks to J. Huchra, M. Kurtz, R. Olowin,
and G. Wegner for sharing their expertise in morphological
classification, and to J. Huchra and A. Mahdavi for each kindly providing
a redshift in advance of publication.  This research was supported in
part by the Smithsonian Institution.
\end{section}

\clearpage
\dummytable
\begin{deluxetable}{lccrccrr@{$\,\pm\,$}l}
\tablecaption{
Imaging and Spectoscopic Survey Targets and Properties\label{brighttab}}
\tablewidth{0pt}
\tablenum{1}
\tablehead{
\colhead{Name} 
& R.A.
& Decl.
& \colhead{$cz$}
& \colhead{Density\tablenotemark{a}}
& \colhead{$r_{SE}$\tablenotemark{b}}
& \colhead{$T$\tablenotemark{c}}
& \multicolumn{2}{c}{EW(H$\alpha$)}\\
& (B1950.0)
& (B1950.0)
& \colhead{(km/s)}
& \colhead{$(n/\bar n)$}
& \colhead{(mag)}
& 
& \multicolumn{2}{c}{(\AA)}
}
\small
\startdata
\multicolumn{9}{c}{\it CfA2 Survey Galaxies} \\
IC 5378       & 00 00 03.98 & $+$16 21 56.9 &  6554 &
1.19 & 13.43 & $-$5 &  $-$0.5 &  0.4\\
00012$+$1555  & 00 01 10.32 & $+$15 54 30.2 &  6636 &
1.14 & 14.84 & $-$3 &  34.2 &  1.9\\
00017$+$1030  & 00 01 39.12 & $+$10 30 42.5 &  8134 &
0.91 & 13.66 & 3 &   8.8 &  1.1\\
NGC 7825      & 00 02 32.74 & $+$04 55 31.1 &  8220 &
1.45 & 13.81 & 3 &  $-$0.1 &  0.6\\
00055$+$0926  & 00 05 32.74 & $+$09 26 22.6 &  6606 &
1.18 & 13.29 & 3 &   6.5 &  0.6\\
00059$+$0956  & 00 05 54.05 & $+$09 55 37.9 &  6608 &
1.16 & 14.06 & $-$2 &   4.1 &  1.2\\
\multicolumn{9}{l}{\nodata}\\
\multicolumn{9}{c}{\it 15R Survey Galaxies} \\
464.015505    & 00 01 02.21 & $+$10 19 30.4 &  8085 &
0.91 & 14.18 & $-$5 &  $-$0.4 &  0.8\\
464.040541    & 00 05 19.08 & $+$11 34 57.7 &  6685 &
1.06 & 15.14 & 5 &   5.4 &  1.8\\
464.069313    & 00 11 08.35 & $+$12 55 00.1 &  8059 &
0.77 & 15.46 & 0 &  10.8 &  3.5\\
465.011685    & 00 26 35.78 & $+$10 19 10.9 &  9887 &
1.78 & 15.21 & 10 &  88.0 &  7.2\\
467.045084    & 01 11 34.99 & $+$12 24 11.5 &  5894 &
1.18 & 15.05 & 3 &  23.2 &  2.5\\
467.017701    & 01 19 00.91 & $+$10 36 51.8 &  9989 &
1.99 & 14.66 & $-$5 &   0.7 &  1.1\\
\multicolumn{9}{l}{\nodata}\\
\multicolumn{9}{c}{\it Century Survey Galaxies}\\
c20.CJ        & 13 33 32.59 & $+$29 28 09.1 &  6414 &
1.38 & 16.48 & 5 & \nodata & \nodata\\
c14.FB        & 13 56 20.90 & $+$29 37 54.1 &  5748 &
0.65 & 16.07 & 10 & \nodata & \nodata\\
c14.HQ        & 14 02 20.71 & $+$29 24 50.0 &  7624 &
1.51 & 15.36 & 3 & \nodata & \nodata\\
c14.JA        & 14 06 04.70 & $+$29 15 11.9 &  7412 &
1.01 & 15.81 & 6 & \nodata & \nodata\\
c14.JK        & 14 07 58.01 & $+$29 40 08.0 &  6574 &
0.38 & 15.76 & 10 & \nodata & \nodata\\
e1390.CX      & 14 44 36.60 & $+$29 23 06.0 &  8316 &
1.21 & 15.49 & $-$2 & \nodata & \nodata\\
\multicolumn{9}{l}{\nodata}\\
\tableline
\enddata
\footnotesize
\tablecomments{
\baselineskip 12pt 
Right ascension in hours, minutes, and seconds of time.  Declination
in degrees, minutes, and seconds of arc.  Radial velocities are Galactocentric.
Table \ref{brighttab} is available in its entirety by request from the authors.
A portion is shown here for guidance regarding its form and content.}
\tablenotetext{a}{Density uncertainty $\lesssim 0.1$ (Grogin \& Geller 1998).}
\tablenotetext{b}{SExtractor MAG\_BEST (cf.~\S2.2).}
\tablenotetext{c}{$B$-band CCD morphology.  Both Irr and Pec assigned $T=10$.} 
\end{deluxetable}
\dummytable
\begin{deluxetable}{ccr@{$\,\pm\,$}lc@{\qquad}lrrl}
\tablecaption{Neighbors to Primary Galaxies in this Study\label{fainttab}}
\tablenum{2}
\tablewidth{0pt}
\tablehead{
\colhead{R.A.}
& \colhead{Decl.}
& \multicolumn{2}{c}{$cz_{\sun}$}
& \colhead{$r_{\rm SE}$\tablenotemark{a}}
& \colhead{Neighbor}
& \colhead{$\Delta\theta$}
& \colhead{$\Delta cz_{\sun}$}
& \\
\colhead{(B1950)}
& \colhead{(B1950)}
& \multicolumn{2}{c}{(km/s)}
& \colhead{(mag)}
& \colhead{Primary\tablenotemark{b}}
& \colhead{($\arcsec$)}
& \colhead{(km/s)}
& \colhead{Comments}
}
\scriptsize
\startdata
00 00 04.01 & $+16$ 22 24.2 & 5834 & 5 & 15.24 & IC 5378 & 27 & $-$515 & NED redshift \\
00 00 06.89 & $+16$ 19 18.5 & 6328 & 18 & 15.33 & IC 5378 & 163 & $-$21 &  \\
00 02 53.21 & $+04$ 53 52.8 & 5629 & \nodata & 14.26 & NGC 7825 & 321 & $-$2426 & NED redshift \\
00 05 14.45 & $+09$ 25 53.4 & \multicolumn{2}{c}{\nodata} & 15.43 & 00055$+$0926 & 272 & \nodata &  \\
00 05 56.50 & $+09$ 56 06.7 & \multicolumn{2}{c}{\nodata} & 15.18 & 00059$+$0956 & 46 & \nodata &  \\
00 17 15.36 & $+06$ 07 39.7 & 27074 & 21 & 15.71 & 00175$+$0612E & 285 & 16937 &  \\
00 17 16.08 & $+06$ 10 54.8 & 10145 & 16 & 15.60 & 00175$+$0612E & 183 & 8 &  \\
00 17 28.03 & $+06$ 11 24.0 & 10360 & 14 & 15.24 & 00175$+$0612E & 15 & 223 &  \\
00 19 53.71 & $+06$ 29 30.1 & 40642 & 45 & 15.78 & 00203$+$0633 & 364 & 33087 &  \\
00 20 31.30 & $+06$ 28 28.2 & 14820 & 29 & 14.07 & 00203$+$0633 & 335 & 7265 &  \\
00 22 40.06 & $+06$ 13 16.3 & \multicolumn{2}{c}{\nodata} & 15.36 & 00227$+$0613 & 71 & \nodata &  \\
00 32 09.12 & $+11$ 57 43.9 & 26870 & 42 & 16.04 & IC 31 & 317 & 17346 &  \\
00 33 48.22 & $+12$ 25 33.2 & 10118 & 26 & 14.52 & 00336$+$1222 & 301 & 54 & 15R redshift \\
00 34 05.21 & $+21$ 19 57.0 & 9310 & 21 & 14.51 & 00341$+$2117 & 152 & 232 &  \\
00 35 09.17 & $+04$ 32 56.8 & \multicolumn{2}{c}{\nodata} & 15.89 & 00353$+$0437 & 361 & \nodata &  \\
00 36 45.17 & $+12$ 50 19.3 & 10618 & 5 & 15.90 & 00367$+$1250S & 70 & 149 & NED redshift \\
00 36 51.46 & $+03$ 44 20.8 & \multicolumn{2}{c}{\nodata} & 15.56 & 00367$+$0341 & 244 & \nodata &  \\
00 36 57.72 & $+20$ 58 12.4 & 17211 & 41 & 16.04 & 00372$+$2056 & 206 & 8063 &  \\
00 36 58.51 & $+03$ 39 32.0 & 4748 & 6 & 15.26 & 00367$+$0341 & 226 & $-$539 & NED redshift \\
00 37 01.90 & $+03$ 44 12.5 & \multicolumn{2}{c}{\nodata} & 15.95 & 00367$+$0341 & 339 & \nodata &  \\
00 37 04.30 & $+20$ 57 36.4 & 17358 & 60 & 14.56 & 00372$+$2056 & 109 & 8210 & NED redshift \\
00 37 20.09 & $+20$ 56 39.8 & 16356 & 40 & 15.81 & 00372$+$2056 & 136 & 7208 &  \\
01 03 31.13 & $+25$ 21 01.4 & 23418 & 40 & 15.54 & 01034$+$2517 & 239 & 16784 &  \\
01 06 42.53 & $+19$ 39 47.2 & \multicolumn{2}{c}{\nodata} & 15.35 & 01068$+$1935 & 248 & \nodata &  \\
01 06 49.42 & $+19$ 33 29.2 & \multicolumn{2}{c}{\nodata} & 15.90 & 01068$+$1935 & 147 & \nodata &  \\
01 08 49.34 & $+01$ 00 47.9 & 6780 & 10 & 15.08 & 01090$+$0104 & 187 & 6 & NED redshift \\
01 11 27.98 & $+12$ 25 00.8 & 16320 & 37 & 15.46 & 467.045084 & 114 & 10571 & 15R redshift \\
01 14 17.62 & $+12$ 44 24.4 & \multicolumn{2}{c}{\nodata} & 15.68 & 01143$+$1245 & 87 & \nodata &  \\
01 14 28.80 & $+12$ 43 25.0 & \multicolumn{2}{c}{\nodata} & 15.66 & 01143$+$1245 & 183 & \nodata &  \\
01 17 26.95 & $+14$ 05 49.2 & \multicolumn{2}{c}{\nodata} & 16.00 & 01174$+$1406 & 54 & \nodata &  \\
01 17 50.93 & $+15$ 48 35.6 & 16143 & 40 & 15.98 & NGC 476 & 246 & 9856 &  \\
01 20 29.09 & $-00$ 35 07.4 & 12987 & 15 & 15.49 & 01206$-$0039 & 274 & 5497 &  \\
01 20 41.42 & $-00$ 35 28.0 & 23112 & 17 & 15.99 & 01206$-$0039 & 199 & 15622 &  \\
01 22 33.26 & $+14$ 36 45.0 & 6376 & 17 & 14.06 & IC 1700 & 166 & 36 & IC 107 \\*
& & \multicolumn{2}{c}{} & & IC 1698 & 176 & $-$188 &  \\
01 22 42.10 & $+14$ 34 43.3 & 6564 & 17 & 13.52 & IC 1700 & 100 & 224 & IC 1698 \\*
& & \multicolumn{2}{c}{} & & IC 107 & 177 & 188 &  \\
01 22 44.62 & $+14$ 36 16.2 & 6340 & 18 & 12.71 & IC 1698 & 99 & $-$224 & IC 1700 \\*
& & \multicolumn{2}{c}{} & & IC 107 & 168 & $-$36 &  \\
\enddata
\footnotesize
\tablecomments{\baselineskip 12pt
Right ascension in hours, minutes, and seconds of time.  Declination in 
degrees, minutes, and seconds of arc.  Radial velocities are heliocentric.
Table \ref{fainttab} is available in its entirety by request from the authors.
A portion is shown here for guidance regarding its form and content.
}
\tablenotetext{a}{SExtractor MAG\_BEST (cf.~\S2.2).}
\tablenotetext{b}{Name as given in Table 1.}
\end{deluxetable}
\begin{table}
\caption{ \label{morphtab}
Coarse Morphological Fraction vs.~Global Density}
\medskip
\begin{tabular}{lcr@{$\pm$}lr@{$\pm$}lr@{$\pm$}l}\tableline \tableline
Sample &  $N$ & \multicolumn{2}{c}{$T<0$} & \multicolumn{2}{c}{$0\leq T\leq 5$} 
& \multicolumn{2}{c}{$T>5$} \\ \tableline
LDVS & 46 & 15 & 5\% & 65 & 7\% & 20 & 6\%\\
HDVS & 104 & 27 & 4\% & 62 & 5\% & 12 & 3\%\\
VPS & 130 & 32 & 4\% & 58 & 4\% & 9 & 3\%\\
\tableline
\end{tabular}
\tablecomments{See Figure~\ref{tridentt} for $T$-type histogram.}
\end{table}
\begin{table}
\caption{ \label{comptab}
Companion Statistics vs.~Global Density}
\medskip
\begin{tabular}{lccccr}\tableline \tableline
& Unpaired & Paired & 
Unknown & $N_{\rm pairs}$ & $\sigma_{\Delta cz}$ \\
Density Sample & & & & & (km/s) \\ \tableline
LDVS \\
\quad $D_p \leq 115h^{-1}$~kpc & 76\% & 15\% & 9\% & 8 & $88\pm22$ \\
\hfill All $D_p$ & 67\% & 22\% & 11\% & 13 & $102\pm20$ \\
HDVS \\
\quad  $D_p \leq 115h^{-1}$~kpc & 68\% & 20\% & 12\% & 30 & $203\pm26$ \\
\hfill  All $D_p$ & 61\% & 28\% & 12\% & 39 & $183\pm21$ \\
VPS \\
\quad  $D_p \leq 115h^{-1}$~kpc & 69\% & 22\% & 9\% & 36 & $266\pm31$ \\
\hfill  All $D_p$ & 63\% & 28\% & 8\% & 64 & $231\pm20$ \\
\tableline
\end{tabular}
\tablecomments{``Unpaired'' galaxies have no other galaxy with $r_{\rm
SE}\leq16.1$ and $\Delta cz\leq1000$~\kms\ within the specified
projected radius $D_p$; ``paired'' galaxies have at least one.  The
``unknown'' fraction are unpaired but with at least one neighbor
within the specified projected radius and lacking a redshift.}
\end{table}
\begin{table}
\caption[]{ \label{ksisotab}
K-S Comparisons of EW(\ha) Distributions vs.~Density}
\medskip
\begin{tabular}{lccccc}\tableline \tableline
& Overall & Unpaired & Paired & Unpaired & Paired \\ 
Density Samples & & \multicolumn{2}{c}{($D_p \leq 115h^{-1}$~kpc)} & 
\multicolumn{2}{c}{(All $D_p$)} \\
\tableline
$P_{\rm KS}$(LDVS vs.~VPS) & 0.43\% & 13\% & 0.073\% & 40\% & 0.084\% \\
$P_{\rm KS}$(LDVS vs.~HDVS) & 2.8\% & 75\% & 0.010\% & 65\% & 1.9\% \\
$P_{\rm KS}$(HDVS vs.~VPS) & 32\% & 4.7\% & 20\% & 31\% & 0.73\% \\
\tableline
\end{tabular}
\tablecomments{``Unpaired'' and ``paired'' as defined in Table 
\ref{comptab}.}
\end{table}
\begin{table}
\caption{ \label{ftesttab}
F-test Comparisons of Pair Velocity Separation vs. Density}
\medskip
\begin{tabular}{lcc}\tableline \tableline
Density Samples & $D_p \leq 115h^{-1}$~kpc & All $D_p$\\ \tableline
$P_{\rm F}$(LDVS vs.~VPS) & 0.59\% & 0.48\% \\
$P_{\rm F}$(LDVS vs.~HDVS) & 3.2\% & 4.4\% \\
$P_{\rm F}$(HDVS vs.~VPS) & 13\% & 12\% \\
\tableline
\end{tabular}
\tablecomments{F-test probabilities of the null hypothesis that two
samples' $\Delta cz$ distributions have the same variance.}
\end{table}
\clearpage
\clearpage
\clearpage
\clearpage
\clearpage
\clearpage
\clearpage
\clearpage
\clearpage
\clearpage
\end{document}